# A Method for Accurate Spatial Focusing Simulation via Numerical Integration and its Application in Optoacoustic Tomography

Maximilian Bader[ab], Philipp Haim[abc], Lukas Imanuel Scheel-Platz[abc], Angelos Karlas[abdef], Hedwig Irl[g], Vasilis Ntziachristos[abh], Dominik Jüstel[abi✉]

[a] Technical University of Munich, Munich, Germany; Chair of Biological Imaging at the Central Institute for Translational Cancer Research (TranslaTUM), School of Medicine and Health.

[b] Helmholtz Zentrum München, Neuherberg, Germany; Institute of Biological and Medical Imaging

[c] Ludwig-Maximilians-Universität München, Munich, Germany.

[d] Technical University of Munich, Munich, Germany; School of Medicine and Health, Clinic and Polyclinic for Vascular and Endovascular Surgery, TUM University Hospital, Klinikum rechts der Isar.

[e] Technical University of Munich, Munich, Germany; Chair for Computer Aided Medical Procedures and Augmented Reality, Department of Informatics.

[f] DZHK (German Centre for Cardiovascular Research), partner site Munich Heart Alliance, Munich, Germany.

[g] Technical University of Munich, Munich, Germany; School of Medicine and Health, Department of Anesthesiology and Intensive Care, TUM University Hospital, Klinikum rechts der Isar.

[h] Technical University of Munich, Munich, Germany; Munich Institute of Robotics and Machine Intelligence (MIRMI).

[i] Helmholtz Zentrum München, Neuherberg, Germany; Institute of Computational Biology.

[✉] Corresponding author. Email: dominik.justel@tum.de, Address: Technical University of Munich, Chair of Biological Imaging, Ismaninger Straße 22, D-81675 Munich, Germany

## Abstract

The spatial sensitivity of an ultrasound transducer, which strongly influences its suitability for different applications, depends on the shape of the transducer surface. Accurate simulation of these spatial effects is important for transducer characterization and design, and for system response modelling in imaging applications.

In optoacoustic imaging, broadband transducers are used to capitalize on the rich frequency content of the signals, but their usage makes highly accurate simulations with general wave equation solvers prohibitively memory- and time-intensive. Therefore, specialized tools for simulating the isolated spatial focusing properties described by the spatial impulse response (SIR) have been developed. However, the challenging numerics of the SIR and the necessity to convolve the SIR with the wave shape generated by the optoacoustic absorber to simulate the system response lead to numerical inaccuracies of SIR-based methods. In addition, the approximation error of these methods cannot be controlled a priori.

To circumvent the problems associated with the explicit calculation of SIR, we propose directly computing the convolution of the required wave shape with the SIR, which we call the spatial pulse response (SPR).

We demonstrate that by utilizing an h-adaptive cubature algorithm, SPR can be computed with significantly higher accuracy than an SIR-based reference method, and the approximation error can be controlled with a tolerance parameter. In addition, the integration of accurate SPR simulations into model-based optoacoustic image reconstruction is shown to improve image contrast and reduce noise artifacts.

Precise system characterization and simulation leads to improved imaging performance, ultimately increasing the value of optoacoustic imaging systems for clinical applications.

# 1. Introduction

Ultrasound (US) transducers are utilized in sonography, and in optoacoustic (OptA; also called photoacoustic) imaging and sensing systems to detect the acoustic waves generated by the respective interaction of ultrasound or light with the investigated target. The size and shape of the transducer surface define a transducer's spatial focusing properties, i.e., the detector's sensitivity depends on the location of the acoustic source.[1] From a systems theory viewpoint, the transducer shape acts as a spatially varying filter on recorded pressure signals. Methods for the computation of this filter face several numerical challenges that limit their accuracy. Enabling accurate simulation of spatial focusing effects is essential for increased fidelity of system characterization and design, and model-based image reconstruction.[2–4]

Two common strategies for simulating spatial focusing of acoustic transducers exist. The first strategy relies on general wave equation solvers like finite elements,[5] boundary elements,[6] or k-space methods[7–9], which can simulate spatial focusing effects in a very general setting. However, for broadband OptA signals, the time- and memory-complexity of general solvers increases steeply with higher accuracy requirements, making simulations infeasible. Therefore, as a second strategy, specialized tools have been developed to approximate the spatial impulse response (SIR), i.e., the isolated effect of a US transducer's spatial focusing, in simplified settings.[10,11] Assuming a homogeneous medium, the SIR describes the detection of a spherical Dirac delta wave in the form of an integral. Analytic solutions of this integral have been derived for flat[12,13] and spherical[14] transducer geometries. For arbitrary transducer shapes, the geometric nature of the integral and the necessity to consider differential effects make the numerics challenging. A common approach to address this challenge is to discretize the transducer into small subelements with known analytic SIR,[15–17] sometimes combined with adaptive refinement of the subelements.[18] However, error estimates are not available for this approach, meaning that a required accuracy cannot be specified a priori.

In addition to the numerical challenges posed by the calculation of SIR itself, a subsequent numerical convolution with the wave shape generated by the OptA source is necessary to simulate OptA pressure signals. If the full system response or the total impulse response (TIR) needs to be simulated, a convolution with the electrical impulse response (EIR) is required. These convolutions are commonly performed at a fixed sampling frequency, potentially adding a source of numerical inaccuracies.[19,20] In conclusion, a method to simulate OptA and US system responses with an input parameter controlling the accuracy is not available.

We hypothesize that we can achieve more accurate simulations of the system response by circumventing the calculation of pure SIR. For that purpose, we introduce the spatial pulse response (SPR), which is defined as the convolution of SIR with a regular temporal pulse (e.g., the OptA N-shape or EIR). SPR is a regular integral that can be directly computed with numerical integration methods. In addition, numerical integration provides parameters to control the approximation error.

We demonstrate that SPR can be computed with an h-adaptive cubature algorithm with significantly higher accuracy compared to an SIR-based reference method, and that the approximation error can be effectively controlled. We also showcase that the increased accuracy of the simulated system response leads to improved performance of model-based OptA tomography reconstructions with increased contrast and substantially decreased noise artifacts, especially out-of-plane artifacts. Ultimately, improved simulations of OptA and US systems facilitate system design and improve imaging performance, supporting clinical translation and application.

## 2. Methods

### 2.1. Modelling and Image Reconstruction

#### 2.1.1. Model of Optoacoustic Wave Detection

We model the OptA generation of US waves and their detection with focused US transducers. While we focus on OptA waves, the results are also applicable to the detection of acoustic waves of different origins by adjusting the source term of the wave equation accordingly. A more detailed account of the theory can be found in a previous publication by our group[21].

We consider the OptA wave equation in thermal and stress confinement, i.e., OptA US waves that are generated by the illumination of tissue with laser pulses with a pulse width much shorter than the thermal and stress relaxation times,

$$\frac{\partial^2}{\partial t^2} p(x,t) - c^2 \Delta p(x,t) = \delta'(t) \Gamma \eta_{th} \mu_a(x) F(x), \qquad (1)$$

where $x \in \mathbb{R}^3$ is the spatial variable, $t \in \mathbb{R}$ is the time variable, $p$ denotes pressure, $c$ is the speed of sound, $\Gamma$ is the Grüneisen parameter, $\eta_{th}$ is the thermal conversion efficiency, $\mu_a$ is the absorption coefficient distribution, $F$ is fluence, and $\delta'$ denotes the derivative of a Dirac delta[22,23]. We assume $c$, $\Gamma$, and $\eta_{th}$ to be constant and that dispersion effects are negligible[24]. The infinitely short source term induces an instantaneous initial pressure $p_0(x) \coloneqq p(x,0) = \Gamma \eta_{th} \mu_a(x) F(x)$ with vanishing initial velocity. Note that $p_0$ is non-negative as a product of non-negative factors.

The solution of eq. (1) for times $t > 0$ is given by

$$p(x,t) = \frac{1}{4\pi c} \frac{\partial}{\partial t} \int_{S_{ct}^2(x)} \frac{p_0(x')}{|x-x'|} dA(x'), \qquad (2)$$

with $S_{ct}^2(x) := \{x' \in \mathbb{R}^3 \mid |x - x'| = ct\}$ being the sphere with a radius $ct$ around $x$, and $dA$ denoting the surface element[8,25].

US transducers accumulate pressure on the transducer aperture $E \subseteq \mathbb{R}^3$. The spatial focusing properties of the transducer that result from this accumulation of pressure are captured in the SIR, $SIR_x^E$, that describes the total pressure on the surface $E$ generated by a spherical delta wave excitation generated at a location $x \in \mathbb{R}^3$ traveling with the speed of sound $c$[10,12,13].

$$SIR_x^E(t) := \int_E \frac{\delta\left(t - \frac{|x - x'|}{c}\right)}{4\pi c |x - x'|} dA(x'). \quad (3)$$

The Dirac delta in this equation is a formal representation of the tempered distribution that integrates a test function over the spherical surface $S_{ct}^2(x)$[21]. The SIR appears naturally when integrating the pressure generated by a point source over the transducer surface[13,26].

The EIR of the transducer summarizes all effects of the system after the cumulation of pressure. Since the system is assumed to operate in a linear regime and to be time-invariant, the effect of the EIR is fully captured by the convolution of the cumulated pressure signal with a function $EIR^E$. The EIR has units of voltage per force and time, since convolution with the EIR converts a temporal force signal into a temporal voltage signal.

In summary, the signal $s_{p_0}^E$ recorded at the transducer $E$ is given by

$$s_{p_0}^E(t) := \int_{\mathbb{R}^3} p_0(x) \delta' * SIR_x^E * EIR^E(t)\, dx, \quad (4)$$

which is sampled with the system's sampling frequency.

We discuss the special form of the pressure wave and the signal for radially symmetric initial pressure, i.e., for situations in which there is a function $p_0^{rad}$ such that $p_0(x) = p_0^{rad}(|x|)$. The solution of eq. (1) then has the form $p(x, t) = p^{rad}(|x|, t)$ for a function $p^{rad}$. Introducing the initial pressure profile $p_0^{profile}(r) := p_0^{rad}(|r|)$ for $r \in \mathbb{R}$, the solution $p^{rad}$ is given by

$$p^{rad}(r, t) = \frac{r - ct}{4\pi r} p_0^{profile}(r - ct), \quad (5)$$

for $t > 0$ and $r > R$, where we assume that $p_0^{rad}(r) = 0$ for $r > R$.[23]

We define the generalized N-shape $N_{p_0}$ of a radial initial pressure distribution as

$$N_{p_0}(t) := 4\pi c r p^{rad}\left(r, \frac{r}{c} + t\right) = -c^2 t p_0^{profile}(ct), \quad (6)$$

i.e., we move with the wave front and rescale the wave form with the traveled distance.

The radial solution (5) can in turn be written as $p^{rad}(r,t) = \frac{1}{4\pi cr} N_{p_0}\left(t - \frac{r}{c}\right)$, i.e., the generalized N-shape describes the wave form of the spherical wave generated by the radial initial pressure. It has units of force per time.

The signal recorded by a transducer for a radial initial pressure located at $x$ is then given by

$$s_{p_0}^E(t) = N_{p_0} * SIR_x^E * EIR^E(t). \qquad (7)$$

Comparing units, we find that the SIR is unitless, since it converts a temporal signal of force per time to a temporal force signal. This agrees with eq. (4), where the term $p_0(x)\delta'(t)$ can be interpreted as the infinitesimal N-shape with units of force per time and volume.

### 2.1.2. Optoacoustic Image Reconstruction

Optoacoustic image reconstruction is the recovery of the initial pressure distribution $p_0$, given the optoacoustic signals acquired by a transducer array $\{E_k\}_{k=1}^K$, which is a collection of $K \in \mathbb{N}$ transducer elements $E_k$. We perform model-based image reconstruction as introduced in our previous publications.[24,27] Denoting the acquired time samples by $\{t_j\}_{j=1}^T$, the sinogram $S$, i.e., the collection of signals acquired by the array, is given by

$$S_{j,k} \coloneqq s_{p_0}^{E_k}(t_j), \quad j = 1, \dots, T, \quad k = 1, \dots, K. \qquad (8)$$

To recover an approximation of the initial pressure distribution, we need to discretize $p_0$. We choose to discretize it in a shift-invariant space

$$V^2(\psi, L) \coloneqq \left\{ f: x \mapsto \sum_{\ell \in L} d_\ell \psi(x - \ell) \mid (d_\ell)_{\ell \in L} \in \ell^2(\mathbb{R}) \right\}, \qquad (9)$$

where the function $\psi \in L^2(\mathbb{R}^3)$ is called the generator, $L = M\mathbb{Z}^3$ with $M \in GL_3(\mathbb{R})$ is a lattice, and $\ell^2(\mathbb{R}, L) \coloneqq \left\{ (d_\ell)_{\ell \in L} \subset \mathbb{R} \mid \sum_{\ell \in L} |d_\ell|^2 < \infty \right\}$ denotes the space of all square integrable sequences indexed by the lattice $L$. In other words, the initial pressure is represented as a superposition of scaled versions of the generator on a lattice. A suitable choice of generator makes the shift-invariant space a closed subspace of $L^2(\mathbb{R}^3)$. The theory of shift-invariant spaces generalizes the theory of sampling in band-limited spaces (Nyquist-Shannon sampling). Choosing the generator to be the sinc-function of a regular lattice $L$, the resulting shift-invariant space is the corresponding band-limited space[28,29].

We select a regular lattice $L_a$ with spacing $a > 0$, i.e., $M = aI_3$, where $I_3$ is the identity matrix, and a Gaussian generator $G_\sigma(x) \coloneqq \frac{1}{\sqrt{(2\pi)^3 \sigma^2}} e^{-|x|^2/2\sigma^2}$ with $\sigma = \frac{a}{2}$. I.e., we work in the space $V^2(G_\sigma, L_a)$. The initial pressure density $p_0$ is approximated by the projection $p_0^{V^2}$ into this space as

$$p_0(x) \approx p_0^{V^2}(x) := \sum_{\ell \in \Omega(L_a)} p_{0,\ell} G_\sigma(x - \ell), \tag{10}$$

where the coefficients $p_{0,\ell}$ are obtained via the so-called prefilter $G_\sigma^{pre}$ via

$$p_{0,\ell} := p_0 * G_\sigma^{pre}(\ell), \quad G_\sigma^{pre}(x) := \mathfrak{F}^{-1}\left[\frac{\overline{\widehat{G_\sigma}}}{\sum_{\ell^\perp \in L_a^\perp} |\widehat{G_\sigma}(\cdot - \ell^\perp)|^2}\right](x), \tag{11}$$

and $\Omega(L_a) \subset L_a$ is a finite subset of $L_a$ with $|\Omega(L_a)| = P \in \mathbb{N}$ that is determined via the sensitivity field of the system.

Here, $\mathfrak{F}[f](k) := \hat{f}(k) := \int_{\mathbb{R}^3} f(x) e^{-ik \cdot x} dx$ denotes the Fourier transform, $\mathfrak{F}^{-1}[f](x) := \frac{1}{(2\pi)^3} \int_{\mathbb{R}^3} f(k) e^{ik \cdot x} dk$ denotes the inverse Fourier transform, $L^\perp := 2\pi M^{-T} \mathbb{Z}^3$ denotes the reciprocal lattice of a lattice $L = M\mathbb{Z}^3$, and $\bar{z}$ denotes complex conjugation of a complex number $z \in \mathbb{C}$.

Since our generator $G_\sigma$ is radially symmetric, the sinogram data is modeled by

$$S_{j,k} = \sum_{\ell \in \Omega(L_a)} p_{0,\ell} N_{G_\sigma} * SIR_\ell^{E_k} * EIR^{E_k}(t_j). \tag{12}$$

According to eq. (6), the Gaussian N-shape $N_{G_\sigma}$ is given by

$$N_{G_\sigma}(t) = -\frac{c^2 t}{\sqrt{(2\pi)^3 \sigma^2}} e^{-(ct)^2/2\sigma^2}. \tag{13}$$

We introduce the linear model of our system $Sys: p_0 \to S$ that maps an initial pressure distribution $p_0$ to a sinogram $S = Sys[p_0]$ via eqs. (10)-(13). We integrated an additional band-pass filter (Butterworth, $10 kHz < f < 12 MHz$) into the model to disregard unreliable frequency bands. Image reconstruction is performed variationally by solving the optimization problem

$$p_0^{rec} := \arg\min_{p_0 \geq 0} \|Sys[p_0] - S^{meas}\|_2^2, \tag{14}$$

where $S^{meas}$ is the measured sinogram. We choose to regularize by early stopping.

## 2.2. The Spatial Pulse Response (SPR)

To simulate optoacoustic signals acquired by a transducer, according to eq. (12), the SIR needs to be computed for every point in in the set $\Omega(L_a)$. While analytic solutions of the corresponding integral are available for certain flat and spherical transducers, [12–14] numerical methods are required for the calculation of the SIR of transducers of arbitrary shape. The geometric nature of the integral that requires the calculation of the intersection of the spherical wave with the transducer element combined with the differential effects of the distributional integrand make the numerics challenging.

We observe that the calculation of the pure SIR is rarely necessary in applications, since the SIR is convolved with different functions, like the N-shape or the EIR, and any real system records samples of a low-pass filtered signal. Therefore, we propose the spatial pulse response (SPR) that is defined as the convolution of the SIR with a function $w$ in time, which we call the pulse

$$SPR_x^{E,w}(t) := w * SIR_x^E(t) := \int_E \frac{w\left(t - \frac{|x - x'|}{c}\right)}{4\pi c |x - x'|} dA(x'). \tag{15}$$

For smooth pulses like the EIR or the Gaussian N-shape, the integrand is a smooth function. Together with regularity assumptions on the transducer surface $E$, the SPR becomes a simple surface integral for each timepoint $t$ and can be solved with numerical integration methods. The ability to evaluate the integral at arbitrary timepoints makes this approach to SPR computation independent of the sampling frequency.

If the generalized N-shape of a radial initial pressure distribution is used as the pulse, the SPR has units of force, since it is the total force exerted on the transducer surface by the pressure wave. Table 1 lists applications of the SPR for different choices of pulse $w$.

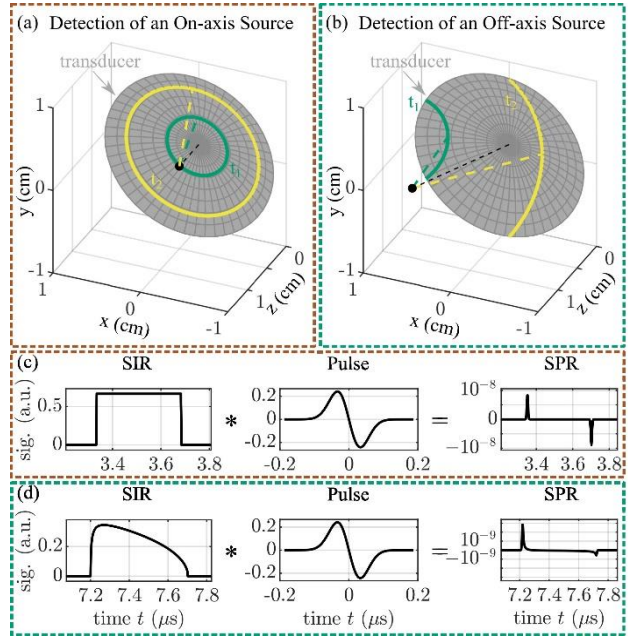

**Figure 1. Spatial Impulse Response (SIR) and Spatial Pulse Response (SPR) to model the wave detection of on- and off-axis sources with a focused transducer**. (a,b) Visualization of the intersection of the spherical pressure wave and the spherical transducer aperture (focal radius: 0.02m, aperture radius: 0.01m) for two point sources (a – on-axis, b – off-axis) and two time points $t_1$ (green line) and $t_2$ (yellow line). (c,d) The SIR convolved with a pulse (a Gaussian N-shape), yields the SPR for the on- and off-axis scenarios depicted in (a) and (b), respectively.

Fig. 1 depicts the process of acoustic wave detection with a spherically focused transducer (focal radius: 0.02m, aperture radius: 0.01m). Fig. 1a,b visualizes the intersection of the transducer aperture with the spherical wave fronts for two point sources – one on-axis (Fig. 1a), and one off-axis (Fig. 1b) – at two different times $t_1$ (green line) and $t_2$ (yellow line). Fig. 1c,d display the SPR as the convolution of the SIR of the spherically focused transducer and the pulse (a Gaussian N-shape) for the source locations in Fig. 1a,b assuming radially symmetric absorbers with a Gaussian profile with $\sigma = 50\mu m$ and a speed of sound of $c = 1500 m/s$.[23]

**Table 1. Applications of the Spatial Pulse Response (SPR).**

| Pulse $w$ | Spatial Pulse Response $SPR^{E,w}$ | Challenges and Applications |
|---|---|---|
| $\delta$ | pure Spatial Impulse Response (SIR) | numerically challenging due to irregular integrand |
| $t \mapsto e^{-itk},\ k \in \mathbb{R}$<br>$t \mapsto \cos(kt),\ k \in \mathbb{R}$ | SIR at frequency $k \in \mathbb{R}$ | computationally expensive for high frequencies |
| sinc | exact values of the projection of the SIR into the corresponding band-limited space at the time samples defined by the sinc | computationally expensive due to the slow decay of sinc |
| Band-pass filter | SIR in the corresponding frequency band | spatial transducer sensitivity in a frequency band |
| Electrical Impulse Response (EIR) | system response to an incoming $\delta$-wave | Total Impulse Response system characterization independent of the incoming waveform |
| N-shape of a radially symmetric absorber | optoacoustic response of the radial absorber | model for the optoacoustic response of a known absorber; important for TIR system characterization |
| N-shape of a radially symmetric basis function of the spatial discretization | optoacoustic response of the radial basis function | Important for numerical simulation of the response of general absorber distributions; applied in model-based image reconstruction |
| N-shape∗EIR | full system response of a radial absorber / basis function | Important for system characterization / numerical simulation of the system |

### 2.2.1. Proposed SPR Implementation – Numerical Integration

We propose the utilization of numerical integration for calculating the SPR directly without the explicit calculation of the SIR. For that purpose, the transducer surface is assumed to be a differentiable two-dimensional manifold. For simplicity, and because no more general case is considered in this paper, we assume that the manifold is parametrized by a single chart $\varphi: E_k \to U \subset \mathbb{R}^2$. The SPR integral is then computed via an h-adaptive quadrature rule that selects optimal weights $a_m \in \mathbb{R}$ and points $u_m \in U$ for evaluating the integrand based on a given required relative tolerance $h_{rel}$. In addition, an absolute tolerance $h_{abs}$ can be specified to avoid unnecessary effort to approximate very low values. The algorithm estimates the error $\epsilon_{est}$ of the current value of the integral by the difference between the values of the integral obtained with a 4th and a 5th order rule using the current set of points $u_m$. The algorithm stops when $\epsilon_{est} < \max(h_{abs}, I \cdot h_{rel})$, where $I \in \mathbb{R}$ is the current absolute value of the integral.[30] The resulting SPR estimate can be formally written as

$$SPR_x^{E_k,w,ours}(t_j) := \sum_{m=1}^{M} a_m f_{x,t_j}^{E_k,w}(u_m), \qquad (16)$$

where the function $f_{x,t_j}^{E_k,w}$ is defined by

$$f_{x,t_j}^{E_k,w}(u) := \frac{w\left(t - \frac{|x - \varphi^{-1}(u)|}{c}\right)}{4\pi c |x - \varphi^{-1}(u)|} \sqrt{\det(D\varphi^{-1})_u^T (D\varphi^{-1})_u}, \qquad (17)$$

with $D$ denoting the Jacobian, and $\det$ the determinant. Note that $SPR_x^{E_k,w}(t) = \int_U f_{x,t}^{E_k,w}(u)\, du$.

The source code is provided at https://github.com/juestellab/spr_jl.

**2.2.2. Reference SIR-based SPR Implementation – Transducer Surface discretization and Numerical Convolution**

We compare the performance of the proposed SPR method with a common reference method[16,17] that relies on the approximation of the transducer surface via simpler subelements with known analytic expressions for the SIR and subsequent numerical convolution with the pulse $w$.

The transducer elements $E_k$ are approximated by a disjoint union of flat rectangular subelements $E_{k,q}$,

$$E_k \approx E_k^{ref} := \bigcup_{q=1}^{Q} E_{k,q}, \quad E_{k,q_1} \cap E_{k,q_2} = \emptyset, \ q_1 \neq q_2, \qquad (18)$$

such that the SIR is approximated by

$$SIR_x^{E_k}(t) \approx SIR_x^{E_k,ref}(t) := \sum_{q=1}^{Q} SIR_x^{E_{k,q},ana}(t), \qquad (19)$$

where $SIR_x^{E_{k,q},ana}$ denotes the analytic SIR for the subelement $E_{k,q}$ and source location $x$.

The size of the subelements determines the approximation accuracy and is usually chosen via the definition of the far-field regime by requiring the diameter $diam$ of the subelements to satisfy $diam \ll \sqrt{\frac{4\,dist_{min}\,c}{f_s}}$, where $dist_{min}$ is the minimal distance between the points in $\Omega(L_a)$ and the element $E_k$, and $f_s$ is the sampling frequency.[10] While this method allows the decrease of the diameter of the subelements and thereby improves the approximation, we note that the quality of the approximation by flat subelements solely depends on the curvature of the element. We introduce a constant $c_\ll \ll 1$ to ensure that the diameters are small enough and an oversampling factor $k_{over}$ to observe the improvement of the approximation at the cost of a higher sampling frequency, i.e., the diameter of the subelements is given by

$$diam(c_\ll, k_{over}) := c_\ll \sqrt{\frac{4\,dist_{min}\,c}{k_{over}f_s}}. \qquad (20)$$

The reference SIR was implemented with the software package Field II.[10,15]

The reference SIR at the chosen sampling frequency was subsequently numerically convolved with the pulse $w$ sampled at the same frequency. We denote this discrete convolution by $*_d$ to clearly differentiate it from continuous convolution. It is defined by $f *_d g(t_j) := \sum_m f(t_m)g(t_j - t_m)$. Finally, the reference SPR $SPR_x^{E_k,ref}$ was obtained by down sampling it to the required sampling frequency or by linearly interpolating and resampling it. Denoting the interpolation operator by $LinInt$, we can write

$$SPR_x^{E_k,w,ref}(t_j) := LinInt\left[w *_d SIR_x^{E_k,ref}\right](t_j). \qquad (21)$$

Note that for down sampling to a subset of samples, the linear interpolation operation is the identity.

### 2.2.3. Ground Truth SPR

To evaluate the performance of our SPR implementation against the reference SPR method, we determined their accuracies against a ground truth SPR, which was obtained by considering a transducer for which the analytical SIR is known [14]. This analytical solution was then convolved with the analytic pulse by using adaptive Gauss-Kronrod integration. To make sure that the accuracy of the convolution was high enough to be considered a ground truth, we used a relative error tolerance of $h_{rel} = 10^{-10}$, and an absolute error tolerance of $h_{abs} = \max(SPR) \cdot 10^{-4} \cdot h_{rel}$. The algorithm accordingly selected time points $t_m$ and weights $b_m$ and computed the SPR via

$$SPR_x^{E_k,w,GT}(t_j) := \sum_{m=1}^{M} b_m \, w(t_m) \, SIR_x^{E_k,ana}(t_j - t_m). \qquad (22)$$

## 2.3. Validation

### 2.3.1. Validation of Accuracy and Efficiency

We compared SPR computation with our method to the reference method relative to the ground truth SPR by comparing the accuracy and efficiency for a spherical transducer $E_{sph}$ (focal radius: $0.02m$, aperture radius: $0.01m$). Radially symmetric initial pressure distributions with Gaussian profiles (standard deviation (std): $50\mu m$) were simulated at 1280 locations ((axial distance, lateral distance) $\in \{(0.005m + j \cdot 0.0005m, k \cdot 0.0005m), j = 0, ..., 60, k = 0, ..., 20\} \setminus \{(0.020m, 0m)\}$). The focal point of the transducer was excluded due to the irregularity of the analytic SIR solution, which reduces to a scaled delta distribution.[14]

Our SPR simulation was performed at a constant sampling frequency $f_s = 40 MHz$. The numerical integration was initialized with 30 regular sampling points in the chosen coordinates. The absolute error tolerance was set to $h_{abs} = \max(SPR) \cdot 10^{-4} \cdot h_{rel}$, and varying relative error tolerances $h_{rel} \in \{10^{-2}, 10^{-3}, 10^{-4}, 10^{-5}, 10^{-6}, 10^{-7}, 10^{-8}\}$ were used to adjust simulation accuracy (see Section B1.).

For the implementation of the reference SPR, we employed the software package Field II,[10,15] and convolved the resulting SIR with the generalized N-shape of the absorber numerically (see Section B2.). To adjust simulation accuracy, we oversampled the SIR by a factor $k = \frac{f_{s,sim}}{40MHz} \in \{1, 2, 10, 25, 125, 250\}$. Since the sampling points at 40 MHz were a subset of the sampling points at higher frequencies, the linear interpolation in eq. (21) reduced to the identity for this experiment. To comply with the far-field regime approximation, we selected $c_{\ll} = 1/10$.

For points on the transducer axis, the ground truth SPR was derived analytically, because the SIR is a rectangular pulse. For points off the transducer axis, we convolved SIR and pulse with adaptive Gauss-Kronrod integration (see Section B3.).

The accuracy was quantified with the root square error $RSE$

$$RSE_{mthd}(x_m) := \left( \frac{1}{f_s} \sum_{j=1}^{T} \left| SPR_{x_m}^{E_{sph},N_{G_\sigma},mthd}(t_j) - SPR_{x_m}^{E_{sph},N_{G_\sigma},GT}(t_j) \right|^2 \right)^{1/2}, \qquad (23)$$

where $mthd \in \{ref, ours\}$.

The efficiency was determined via the computation time $t_{mthd}^{comp}(x_m)$ on a personal computer with an AMD Ryzen Threadripper 3970X processor (32 cores, 3.7GHz base speed), 256GB RAM and Windows 10 operating system.

In the Results section, we report median RSEs and computation times across the 1280 locations, i.e., median accuracies and efficiencies, for given oversampling factors and relative tolerances for the reference method and our method, respectively.

### 2.3.2. Validation in the Image Domain

We validated the effect of the improved accuracy of the SPR simulation on the image quality. For that purpose, we integrated our SPR and the reference SPR with the Gaussian N-shape as the pulse in the system model for model-based image reconstruction. More precisely, we used the methodology of the individual synthetic total impulse response (isTIR) [24]. Measurements of the optoacoustic signals generated by small homogeneous spherical absorbers at locations $x_m$ were used together with simulations of the SPR to infer the EIR of the individual elements $E_k$ by solving the linear system

$$s_{p_m}^{E_k}(t_j) = SPR_{x_m}^{E_k, N_{hom}} * EIR^{E_k}, \tag{24}$$

where $p_m$ is the radial initial pressure generated in the homogeneous absorber at location $x_m$, assuming that it is homogeneously illuminated, and $N_{hom}$ is the N-shape of a homogeneous absorber

$$N_{hom}(t) = -c^2 t \, rect\left(\frac{ct}{R}\right) \tag{25}$$

with radius $R > 0$, where $rect(t) = 1$ for $|t| \leq \frac{1}{2}$, and $rect(t) = 0$ otherwise. We call the EIR derived via the respective SPR implementation $EIR_{mthd}^{E_k}$ with $mthd \in \{ref, ours\}$. Accordingly, the system response implemented with the respective signal model is called $Sys_{mthd}$ and the reconstructed initial pressure obtained via the solution of the optimization problem (14) is called $p_0^{mthd}$.

The optimization problem in (14) was solved with an iterative scheme using a non-negative least squares solver.[24,31] The reconstruction was regularized by early stopping after 50 outer and 2 inner iterations.[32]

The image quality achieved by the two implementations of SPR was assessed with several metrics. The relative residual norm

$$MSE_{rel}^{mthd} = \frac{\sum_{k=1}^{K} \sum_{j=1}^{T} \left| Sys_{mthd}[p_0^{mthd}]_{j,k} - S_{j,k}^{meas} \right|^2}{\sum_{k=1}^{K} \sum_{j=1}^{T} |S_{j,k}^{meas}|^2}, \quad (26)$$

was used to assess the ability of the respective model to explain the measured sinogram data.

Contrast resolution (CR) was used to quantify image contrast. CR was obtained by segmenting image regions with strong absorbers (vessels) and background regions in $p_0^{mthd}$ and collecting the corresponding pixel values in the sets $\Omega_{abs}^{mthd}$, and $\Omega_{back}^{mthd}$, respectively. CR was then defined by

$$CR^{mthd} = \frac{\frac{1}{|\Omega_{abs}^{mthd}|} \sum_{p_{abs} \in \Omega_{abs}^{mthd}} |p_{abs}|^2 - \frac{1}{|\Omega_{back}^{mthd}|} \sum_{p \in \Omega_{back}^{mthd}} |p_{back}|^2}{\frac{1}{|\Omega_{abs}^{mthd}|} \sum_{p_{abs} \in \Omega_{abs}^{mthd}} |p_{abs}|^2 + \frac{1}{|\Omega_{back}^{mthd}|} \sum_{p \in \Omega_{back}^{mthd}} |p_{back}|^2}, \quad (27)$$

where |Ω| denotes the number of elements of a set Ω.

As an additional image quality metric, we estimated the noise std ($\sigma_{noise}$) in an image utilizing a robust noise estimator implemented in the Python Scikit-Image package.[33] The algorithm estimates the noise using the median absolute deviation of the wavelet coefficients at the most detailed level. The wavelet coefficients at fine levels empirically contain mostly noise and possible bias by a signal is controlled by analyzing the median absolute deviation.[34]

We analyzed different images obtained from the reconstructed MSOT data: in-plane (x-z plane) central cross-sectional images, out-of-the-plane MIPs along the out-of-plane y-axis, and MIPs along the lateral x-axis.

As a validation dataset, we selected MSOT scans, acquired from three healthy volunteers proximal to their elbows, which have been presented in a previous publication.[35] Informed consent was received from all volunteers. The ethics committee of the Technical University of Munich approved the publication of the data. The study protocol is described in detail in the initial publication.

The images were acquired with an MSOT Acuity Echo Prototype (iThera Medical GmH), a clinical hand-held hybrid Multispectral Optoacoustic and Ultrasound Imaging (MS-OPUS) system. Regular ultrasound gel was used to couple optoacoustic waves generated in tissue into the probe. The handheld MS-OPUS system consisted of a one-dimensional array (256 elements, radius: 6cm, angular coverage: 145°) of concave focused transducers (elevational radius: 6.5cm, elevational arc length: 26mm, width: 0.49mm, center frequency: 4.3 MHz, 6dB-bandwidth: 50%).

In preprocessing, the signal data was denoised with a deep learning-based algorithm to remove electrical noise artifacts[31] and afterwards, the signals were band pass filtered (Butterworth, $10k\text{Hz} < f < 12M\text{Hz}$).

## 3. Results

### 3.1. A Numerically Accurate and Efficient Simulation of Optoacoustic Wave Detection for a Focused Transducer

Our method to calculate SPR allows simulation of OptA wave detection with a focused transducer at a higher numerical accuracy than the SIR-based reference method. In addition, the error tolerance input provided by the method allows a priori control of the simulation accuracy.

Fig. 2a displays the simulation scenario with a spherically focused transducer (focal radius: $0.02m$, aperture radius: $0.01m$) and radially symmetric sources with Gaussian profiles (std $\sigma = 50\mu m$) arranged in a two-dimensional grid along the lateral and focal axes.

Fig. 2b shows the median RSE over the median computation time (median over the 1280 different simulated sources) for the simulated SPR using our method $SPR^{ours}$ (orange) and the SIR-based reference method $SPR^{ref}$ (black) relative to the ground truth SPR (see Methods 3.2 and 3.3.1, for implementation details). The RSE of the methods was evaluated at a fixed sampling frequency $f_s = 40MHz$. By reducing the relative tolerance $h_{rel}$ and the oversampling factor $k$, the median RSEs of the proposed and the reference method decrease, while the computation time rises (Fig. 2b). However, while the RSE for the reference SPR stagnates, we demonstrate that our proposed method allows us to approximate the SPR with a median RSE that is up to five orders of magnitude lower than the reference method with comparable computation time (Fig. 2b). The median error estimate $\epsilon_{est}$ (blue) is higher than the median RSE, demonstrating the ability to control the error with the relative tolerance parameter $h_{rel}$ (Fig. 2b).

The RSEs for the individual simulated sources for three values of the parameters $h_{rel}$ and $k$, are shown in Fig. 2c and Fig. 2d for the proposed and the reference method, respectively. The examples were chosen, so that the median computation time is comparable column-wise. Although the median RSE is similar for both methods in the leftmost column in Fig. 2c and Fig. 2d, our method provides RSEs several orders of magnitude lower than the reference method close to the focus and along the axis. While the RSEs improve uniformly in the whole simulation domain for the proposed method from left to right, the RSE stagnates for the reference method.

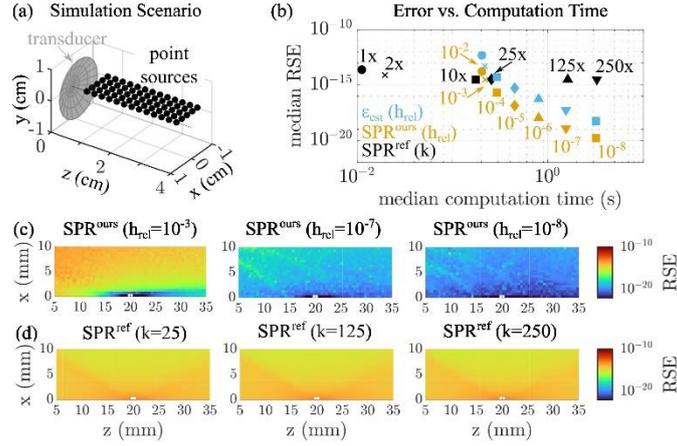

**Figure 2: The proposed Spatial Pulse Response (SPR) computation ($SPR^{ours}$) simulates optoacoustic wave detection more accurately than the SIR-based reference method ($SPR^{ref}$) for a spherically focused transducer and allows a priori control of the accuracy.** (a) Schematic depicting the simulation scenarios consisting of the spherical transducer (focal radius: $0.02m$, aperture radius: $0.01m$) and the various optoacoustic sources (radially symmetric Gaussian absorber, $\sigma = 50\mu m$) distributed in the transducer's field of view. (b) Comparison of median root squared error (RSE) over median computation time for all points in the field of view using $SPR^{ours}$ (orange) and $SPR^{ref}$ (black) compared to the ground truth SPR. The median error estimate $\epsilon_{est}$ is displayed in blue. The simulation accuracy is tuned by the relative tolerance ($h_{rel}$) and the oversampling factor ($k$) for the $SPR^{ours}$ and $SPR^{ref}$, respectively. (c) The simulation accuracy for the individual simulated sources for the $SPR^{ours}$ with three values of $h_{rel}$. (d) The simulation accuracy of the individual simulated sources for $SPR^{ref}$ with three values of $k$, chosen so that computation times are comparable with the panel from (c) directly above.

## 3.2. Accurate Simulation of Spatial Focusing Improves Model-Based Image Reconstruction

The high accuracy of the proposed method for simulating the spatial focusing of transducers directly translates into increased fidelity of the system model and reduced noise artifacts in model-based MSOT image reconstructions. This effect is demonstrated in scans of three representative human arms proximal to the elbow.[35] OptA images at 700nm and the co-registered cross-sectional US images are displayed in Fig. 3, showing blood vessels and the morphology of the surrounding tissue, respectively. These three scans will be referred to as Scan 1, 2, and 3 from here on.

Fig. 3a-d showcases images from Scan 1. Cross-sectional OptA images at 700nm reconstructed using the proposed SPR and the reference SPR methods were overlaid on the US image in Fig. 3a and b, respectively. The prominent blood vessels (numbered 1 and 2) qualitatively show higher contrast in Fig. 3a (proposed SPR reconstruction) than in Fig. 3b (reference SPR reconstruction). Moreover, less noise is qualitatively observed in-plane for the proposed SPR reconstruction (Fig. 3a) compared to the reference SPR reconstruction (Fig. 3b). Most prominently, the noise level is considerably decreased in the out-of-plane Maximum Intensity Projections (MIPs) for the proposed SPR method (Fig. 3c, arrow 3) compared to the reference SPR reconstruction method (Fig. 3d arrow 3).

These results (Fig 3a-d) were corroborated by images from Scans 2 (Fig. 3e-h) and 3 (Fig i-l), which were identically arranged. The contrast of the vessels in Scan 2 (number 6) and Scan 3 (number 8) is qualitatively slightly better for the proposed SPR reconstructions in Fig. 3e,i than for the corresponding reference SPR reconstructions in Fig. 3f,j. Stronger noise artifact reductions are observed in the out-of-plane MIPs in Fig. 3g,k using the proposed SPR method than using the reference SPR method in the corresponding MIPs in Fig. 3h,l. Clear noise artifacts that are visible in the reference reconstructions (Fig. 3h arrow 7 and Fig 3l arrow 9) are removed with the proposed SPR reconstruction (Fig. 3 g,k).

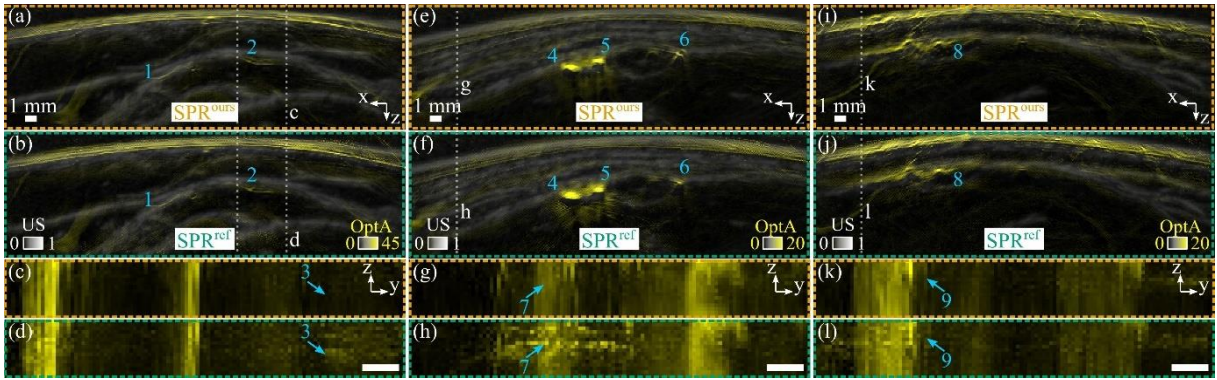

**Figure 3: Multispectral Optoacoustic Tomography (MSOT) images of three human forearms reconstructed with the proposed SPR method display slight contrast enhancement, and considerable noise reduction in-plane and out-of-plane.** (a,b) Overlaid cross-sectional ultrasound (gray) and optoacoustic (700nm, yellow) images (x-z-plane) from the first human forearm (Scan 1) reconstructed with the proposed SPR (outlined in brown) and the reference SPR (outlined in green) method, respectively. Two prominent blood vessels (number 1 and 2) showcase qualitatively higher contrast in the reconstruction with the proposed compared to the reference SPR method. (c,d) Out-of-plane maximum intensity projections (MIPs) along the lateral x-axis (x-MIPs) using (c) the proposed SPR method and (d) the reference SPR method. The shown x-MIPS are from the respective regions of interest delineated in (a) or (b) by vertical grey dotted lines. Noise artifacts near the bottom of the image reconstructed with the reference SPR method (arrow 3) are not seen when using the proposed SPR method. (e-h) Overlaid cross-sectional ultrasound and optoacoustic images, and out-of-plane MIPs from Scan 2, reconstructed with the proposed SPR and the reference SPR method identically arranged as in (a-d). Numbers 4-6 display blood vessels with high contrast and arrow 7 out-of-plane noise. (i-l) Overlaid cross-sectional ultrasound and optoacoustic images and out-of-plane MIPs from Scan 3 with the proposed SPR and the reference SPR method identically arranged as in (a-d). Numbers 8 and 9 display a high contrast blood vessel and out-of-plane noise artifacts, respectively. All scalebars have a length of 1mm and the colormap is identical for all images from each scan.

Table II shows the quantitative analysis of the qualitative observations in Fig. 3, where several metrics for reconstruction quality were compared between in- and out-of-plane images of a scan, and among all three scans. The metrics compared include relative squared residual norm ($MSE_{rel}$), contrast resolution (CR), and estimated noise std ($\sigma_{noise}$) (see Methods 2.3.2).

For Scan 1, all metrics improve with the proposed method to compute SPR, demonstrating improved accuracy and contrast, and reduced noise. The $MSE_{rel}$ of the proposed SPR reconstruction is 2.23% lower compared to the reference SPR reconstruction. The proposed SPR reconstruction results in the 2$^{nd}$ vein (number 2 in Fig. 3) having a 0.71% larger CR, and 0.384 lower $\sigma_{noise}$ in the in-plane image than the reference SPR reconstruction. For the MIP along the out-of-plane axis (y-MIP), $\sigma_{noise}$ is 0.483 higher for the reference method than the proposed method. The discrepancy in $\sigma_{noise}$ between the reference and proposed reconstruction is largest in the x-MIPs with a value of 1.24.

The metrics for Scans 2 and 3 confirm the increased performance of the proposed method. The $MSE_{rel}$ of the MSOT reconstructions are 23.7% and 31.8% lower for Scans 2 and 3, respectively, when using the proposed SPR method instead of the reference method. The CR of the left (number 4) and right vessel (number 5) are 0.72% and 0.05% higher using the proposed method compared to the reference method. Similarly, a prominent vessel in Scan 3 (arrow 8 in Fig. 3i,j) has a CR that is 6.5% higher with the proposed method than the reference method. In images of Scans 2 and 3, $\sigma_{noise}$ is 0.222 and 0.234 smaller in the cross-sectional in-plane image and 0.305 and 0.277 smaller in the y-MIP, respectively, when the proposed method was compared to the reference method. Due to the removal of the noise artifacts, $\sigma_{noise}$ is 0.877 and 0.566 lower in the x-MIP for Scans 2 and 3 using the proposed SPR compared to the reference SPR.

In summary, using the proposed SPR method for system characterization and model-based image reconstruction increased the image reconstruction quality for a hand-held MSOT system. While CR increased only slightly, noise artifacts in in-plane images, and in x- and y-MIPs decreased considerably, improving out-of-plane signal localization.

|  | Scan 1 (Fig.3a-d) | | Scan 2 (Fig.3e-h) | | Scan 3 (Fig.3i-l) | |
| --- | --- | --- | --- | --- | --- | --- |
|  | SPR[ref] | SPR[ours] | SPR[ref] | SPR[ours] | SPR[ref] | SPR[ours] |
| $MSE_{rel}$ | 8.45% | **6.22%** | 31.4% | **7.74%** | 35.6% | **3.78%** |
| CR (blue numbers 2,4,5,8) | 91.3% | **92.0%** | 97.0% 97.8% | **97.7% 97.9%** | 89.8% | **96.3%** |
| $\sigma_{noise}$ in-plane | 1.20 | **0.820** | 0.758 | **0.536** | 0.680 | **0.466** |
| $\sigma_{noise}$ y-MIP out-of-plane | 1.22 | **0.736** | 0.720 | **0.415** | 0.708 | **0.431** |
| $\sigma_{noise}$ x-MIP out-of-plane | 1.61 | **0.367** | 1.16 | **0.283** | 0.850 | **0.284** |

**Table II: Quantitative analysis of *in-vivo* reconstructions using the proposed SPR method ($SPR^{ours}$) and the reference SPR method ($SPR^{ref}$).** Human arm Multispectral Optoacoustic Tomography (MSOT) image reconstructions with the $SPR^{ours}$ (Fig. 3a,c,e,g,i,k) result in a lower relative residual norm ($MSE_{rel}$), higher contrast resolution (CR) of prominent vessels and lower estimated noise standard deviations ($\sigma_{noise}$) in cross-sectional images (in-plane) and maximum intensity projections (MIPs) compared to $SPR^{ref}$ reconstructions (Fig. 3b,d,f,h,j,l). The best values among the two methods are denoted in bold.

## 4. Discussion

We present a numerical solution for accurate simulation of the spatial focusing properties of US transducers combined with the ability to control the approximation error. Our approach avoids the numerical challenges posed by the simulation of pure SIR and subsequent convolution with the absorber response or EIR. The increased accuracy of our solution facilitates reliable system characterization and modeling and improves the performance of OptA image reconstruction.

While it is necessary to separate different concepts, like SIR, EIR, and the N-shape, to better understand the phenomenology of optoacoustics, the sequential computation of the individual parts and their combination can result in numerical problems. We show that the theory of numerical integration can be utilized to circumvent such problems and achieve highly accurate simulation results. A major advantage of numerical integration theory over heuristic methods is the availability of error estimates that allow control of the approximation accuracy according to the simulation requirements. This leads to reliable simulations and system characterizations and models of high quality.

Using the proposed method to calculate SPR for model-based image reconstruction led to improved contrast and markedly reduced noise artifacts compared to the reference method. The reduction of artfacts was especially substantial out-of-plane, allowing image contamination from out-of-plane absorbers to be more reliably identified, which is of great value for clinical applications.[22,23] Another benefit is that the out-of-plane information is a valuable source of information that can be used to combine multiple acquired slices to reconstruct volumetric data, for example, in robotic applications of OptA or US imaging.[24–26]

Due to the flexibility of the SPR concept, it can be used to compute many different objects, including the OptA response of absorbers, the TIR, and SIR in a frequency band (see Table I). The underlying geometric integral of the SIR causes numerical challenges. Such geometric integrals also appear in many other applications, e.g., whenever the Radon transform and its generalization is used. Computational problems might, in general, be avoidable, if the calculation of numerically challenging integrals can be circumvented by reformulating the object that needs to be computed as a regular integral.

The main limitation of the proposed method to calculate SPR integrals is the computation time that can be substantial if system responses in a large field of view need to be simulated at high spatial and temporal resolution. Artificial neural networks have recently been used to implement algorithms for fast evaluation of complex functions[28,29] and might be a solution that is simultaneously fast and highly accurate.

High-quality simulations of system components are required for reliable engineering solutions, especially in the medical imaging field with its strict requirements for high reliability. Methods like the proposed algorithm that allow control of the accuracy according to application requirements, thus, have a direct impact on the clinical translation of medical imaging devices.

## Acknowledgements

This project has received funding from the Bavarian Ministry of Economic Affairs, Energy and Technology (StMWi) (DIE-2106-0005// DIE0161/02, DeepOpus), the Munich Institute of Robotics and Machine Intelligence (MIRMI) of the Technical University of Munich (TUM) (RoMSOT), the Deutsche Forschungsgemeinschaft (DFG) - 455422993 as part of the Research Unit FOR 5298 (iMAGO, subproject TP2, GZ: NT 3/32-1), and the European Research Council (ERC) under the European Union's Horizon Europe research and innovation programme under grant agreement No 101041936 (EchoLux) and under Horizon 2020 under No 694968 (PREMSOT).

The authors would like to thank Antonia Longo, Guillaume Zahnd, Manuel Gehmeyr and Kaushik Basak Chowdhury for the fruitful discussions and Robert Wilson and Serene Lee for their assistance with the preparation of the manuscript.

## Declaration of Competing Interest

Vasilis Ntziachristos is a founder and equity owner of Maurus OY, sThesis GmbH, iThera Medical GmbH, Spear UG and I3 Inc. All other authors declare no competing interests.